\documentstyle[11pt,epsf]{article}
\newcommand\ben{\begin{equation}}
\newcommand\een{\end{equation}}
\newcommand\bea{\begin{eqnarray}}
\newcommand\eea{\end{eqnarray}}

\def\ct#1{$^{\ref{#1}}$}
\def\twoct#1#2{$^{\ref{#1},\ref{#2}}$}
\def\sec#1{\vskip 24pt \noindent {\bf #1} \vskip 12pt}

\def\({\left(}
\def\){\right)} 
\def\[{\left[} 
\def\]{\right]}
\def\pa{\partial}
\def\half{{\mathchoice{{\textstyle{1\over 2}}}{1\over 2}{1\over 2}{1 
\over 2}}}

\def\al{\alpha}
\def\be{\beta}

\def\de{\delta}
\def\ep{\epsilon}

\def\th{\theta}

\def\si{\sigma}

\def\Th{\Theta}

\def\Om{\Omega}

\def\vph{\varphi}

\def\ncs{N_{\rm CS}}
\def\phh{\hat\phi}
\def\Phh{\hat\Phi}
\def\fem{F^{\rm em}}
\def\thw{\th_{\rm W}}
\def\F{{\cal F}^{\rm em}}
\def\bx{\mbox{\boldmath $x$}}
\def\bxh{\hat{\bx}}
\def\bsi{\mbox{\boldmath $\si$}}
\def\bxi{\mbox{\boldmath $\xi$}}
\def\xih{\hat{\xi}}

\def\overleftrightarrow#1{\vbox{\ialign{##\crcr
    $\leftrightarrow$\crcr\noalign{\kern-1pt\nointerlineskip}
    $\hfil\displaystyle{#1}\hfil$\crcr}}}
\def\dbw{\overleftrightarrow\partial}

\begin{document}
\begin{flushright}SUSX-TH-94/71 \\ (May 1994)\\  \end{flushright}
\null\vskip 60pt
\sec{SPHALERONS AND STRINGS}
\vskip 24pt
\begin{quotation}
\noindent{Mark Hindmarsh} \\


\noindent{School of Mathematical and Physical Sciences, \\
University of Sussex, \\ Brighton BN1 9QH\\}
\vskip 12pt
\end{quotation}
 
\sec{INTRODUCTION}  
This work is based on a paper with Margaret
James.\ct{HinJam94}  In it we showed that the dipole moment of the
sphaleron has its origin in two components: a ring of electric current
circulating around the edge of the sphaleron; and also two regions of
opposite magnetic charge above and below the ring.  This magnetic charge
has a partly topological explanation, arising from the fact that the
sphaleron is axisymmetric and parity invariant.  Here, I shall make a
few remarks that are complementary to our paper.  Firstly, I discuss
the definition of the electromagnetic field and its sources in the
Standard Model, comparing ours with
the better-known one of 't Hooft.\ct{tHo74}  Both allow magnetic charges
and currents, but 't Hooft's magnetic charge is designed purely to count
zeroes of the Higgs field and is not a very satisfactory dynamical
quantity. Secondly, I summarize the results of our paper,\ct{HinJam94} 
which uses our definition of the electromagnetic field to
look inside the sphaleron and find the distribution of charges and
currents which give rise to its dipole moment.  Finally, I go into a bit
more detail about the resemblance between the sphaleron and Nambu's
``dumb-bell''\ct{Nam77} -- 
a segment of $Z$-string which connects a monopole to an
antimonopole.  It has been conjectured that the string segment is a
kind of ``stretched'' sphaleron:\twoct{Vac93}{BarVacBuc94} however, it is
possible to estimate the Chern-Simons number $\ncs$ using the topology
of the Higgs field alone, providing the gauge is chosen correctly. In
this gauge it can be shown that the string segment lacks a crucial
twist in its field configuration that in the sphaleron results in $\ncs
= \half$. 

\sec{ELECTROMAGNETISM IN THE STANDARD MODEL}
In the Standard Model there is no unambiguous definition
of the electromagnetic field.  The first step, on which all agree, is to
use a unit isovector 
\ben 
\phh^a= \Phi^{\dag} \si^a \Phi / |\Phi|^2, 
\een
to project out the massless component of the gauge potential:
\ben
A_\mu = -W^a_\mu\phh^a \sin\thw + Y_\mu \cos\thw,
\een
where $W^a_\mu$ and $Y_\mu$ are the SU(2) and U(1) gauge fields
respectively.  
This reduces to the usual
expression when the Higgs takes its conventional vacuum expectation
value $\Phi=(0,1)^{\rm T}v/\sqrt 2$, for then $\phh^a = - \de^a_3$.
We can also project out the massless component of the
gauge field strength tensor: 
\ben 
\fem_{\mu\nu} = -F^a_{\mu\nu}\phh^a\sin\thw + F^0_{\mu\nu}\cos\thw.  
\label{eEMf}
\een 
However, there is a remnant of the non-Abelian theory in which it is
embedded, for the relationship between $\fem_{\mu\nu}$ and $A_\mu$
does not take the simple form of ordinary electromagnetism. In fact,
\ben 
\fem_{\mu\nu} = \pa_{[\mu}A_{\nu]} + {\sin\thw\over g} \[
\phh^a\pa_\mu\phh^b\pa_\nu\phh^c - \phh^aD_\mu\phh^bD_\nu\phh^c\]
\ep^{abc},
\label{eEMfNA}  
\een
where $D_\mu\phh^a = \pa_\mu + g\ep^{abc}W^b_\mu\phh^c$. 
(A sign error in Ref. \ref{HinJam94} has been corrected here.) This means
that the Bianchi identity is not satisfied, and there is a magnetic
current given by the right hand side of 
\ben
\half\ep^{\mu\nu\rho\si}\pa_\nu\fem_{\rho\si} = - \half
\sin\thw\ep^{\mu\nu\rho\si} F^a_{\nu\rho}D_\si\phh^a.
\label{eMagC}
\een
Note that $D_\si\phh^a$ is orthogonal to $\phh^a$, so that the magnetic 
current is comprised of gauge fields orthogonal both to the electromagnetic 
field and to the Z field, $F^{\rm Z}_{\mu\nu} = -F^a_{\mu\nu}\cos\thw 
- F^0{\mu\nu}\sin\thw$.  
In other words, magnetic currents in the Standard Model 
are made of W fields.  W fields also make electric currents, for one 
can show that 
\ben
\pa^\nu\fem_{\mu\nu} = -\sin\thw F^a_{\mu\nu}D^\nu\phh^a.
\label{eEleC}
\een
The Higgs field does not contribute directly to this expression. This is 
to be expected, for the physical Higgs field is neutral.

A definition that many authors use is that of 't Hooft, who gives the 
electromagnetic field as
\ben
\F_{\mu\nu} = \fem_{\mu\nu} + {\sin\thw\over g} 
\phh^aD_\mu\phh^bD_\nu\phh^c \ep^{abc}.
\label{eEMft}
\een
The purpose of adding the extra term is to force the electromagnetic field 
to satisfy the Bianchi identity almost everywhere, for
\ben
\half\ep^{\mu\nu\rho\si}\pa_\nu\F_{\rho\si} = {\sin\thw\over 2g} 
\ep^{\mu\nu\rho\si}\pa_\nu\phh^a\pa_\rho\phh^b\pa_\si\phh^c\ep^{abc}.
\een
The right hand side of this expression vanishes everywhere except along 
world lines around which  
$\phi^a$ takes ``hedgehog'' configurations.\ct{tHo74}  Thus the 
't Hooft magnetic flux out of a closed surface serves merely to count 
zeroes of the Higgs field.

The two expressions (\ref{eEMf}) and (\ref{eEMft}) agree in the Higgs 
vacuum, where $D_\mu\phh^a=0$.  Away from the vacuum they disagree, and   
there is no absolute standard by which to judge them.  However, the first 
definition has the advantage that the energy density in the magnetic 
field is always $\fem_{ij}{\fem}^{ij}$, whereas the 't Hooft magnetic 
field energy density must be corrected with derivatives of the isovector 
$\phh^a$.  Furthermore, there seems no reason to try and satisfy the 
Bianchi identities almost everywhere -- we know that there is magnetic 
charge in non-Abelian theories, so why not have it spread out and 
created out of physical fields of the theory?

Accordingly, we shall use definition (\ref{eEMf}) in what follows.  With 
it, we can look inside the sphaleron and find that at non-zero 
Weinberg angle the W fields do provide both electric currents and magnetic
charges.

\sec{INSIDE THE SPHALERON}
The sphaleron at zero Weinberg angle is a spherically symmetric 
solution of the classical field equations of the 
Standard Model, which can be written in the following 
form, in the radial gauge $W_r = 0$:
\ben
\Phi = U(\bxh)\pmatrix{0\cr1\cr} h(r) \frac{v}{\sqrt 2}, \qquad 
W = -{2i\over g} dUU^{-1}f(r),
\label{eSph}
\een
where $U = i\bxh\cdot\bsi\si_2$.  The isovector field around this 
configuration is
\ben
\phh^a = 2\hat{x}^a\hat{x}^3 - \de^{a3}.
\een
This field 
is illustrated in Figure 1.  In the core of the sphaleron, the fields 
leave the vacuum and there is non-zero $D_i\Phi$ and $F^a_{ij}$.  The 
spherical symmetry of the energy density  is in fact guaranteed by the 
custodial SU(2) symmetry of the Standard Model, so it is understandable 
that when $\thw\ne 0$ the equal-energy contours of the sphaleron 
solution become prolate.\ct{KleKunBri92}  The sphaleron also develops 
a magnetic dipole moment, which has been calculated perturbatively 
in the small $\thw$ limit,\twoct{KliMan84}{Jam92} and also numerically 
for general values of $\thw$.\ct{KleKunBri92}


\begin{figure}
\epsfxsize=2.5in \centerline{\epsfbox{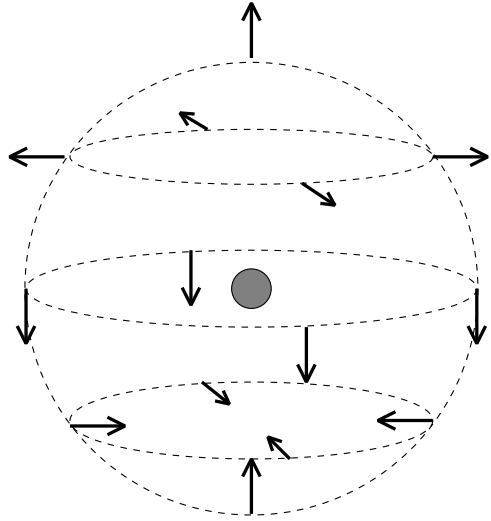}}
\caption
{
Isovector field configuration $\phh^a$ around the sphaleron 
at $\thw=0$.
}
\end{figure}

With the definition (\ref{eEMf}) of the electromagnetic field we can 
understand this dipole moment in terms of magnetic charges and 
electric current defined in a gauge invariant way.  Let us evaluate 
them to first order in $\thw$, where the perturbation to the background 
configuration does not appear.  Using equations (\ref{eMagC}) and 
(\ref{eEleC}) we find that the magnetic charge density 
$\tilde\rho$ and the 
electric current $J_i$ are given by
\ben
\tilde\rho = \frac{g'}{g} 8\cos\th\frac{f'(1-f)}{gr^2}, \qquad
        \mbox{\boldmath$J$}= \frac{g'}{g} 8\sin\th\frac{f(1-f)^2}{gr^4} 
        {\hat{\mbox{\boldmath$e$}}_\varphi},
\een
where ${\hat{\mbox{\boldmath$e$}}_\varphi}$ is the azimuthal unit vector 
in spherical polar coordinates.  The magnetic dipole moment of this 
charge and current distribution is
\ben
\frac{g'}{g}\frac{32\pi}{3g} \int_0^\infty\[rf'(1-f) + f(1-f)^2\]dr.
\een
When the function $f$ is found numerically,\ct{Jam92} we find that the 
charge and the current contribute 70\% and 30\%  respectively to the 
dipole moment.  Moreover, this expression can be shown to be 
{\it identical} to that derived by Klinkhamer and Manton.\ct{KliMan84} 
The total magnetic charge in each hemisphere is $4\pi\thw/g$, which 
is $\thw$ multiplied by the charge of a 't Hooft-Polyakov  
SU(2) monopole\twoct{tHo74}{Pol74}. Indeed, the 
isovector field configuration looks as if it could be the Higgs field 
around a monopole-antimonopole pair in the Georgi-Glashow model.  
This might make one wonder if the presence of the magnetic charge was 
in some way topological.

The topology lies in the subspace of field configurations which are 
restricted to be axisymmetric and parity invariant. 
On the equatorial plane these configurations have 
$\phh^a$ constant, which can be 
chosen to be $-\de^a_3$, and the SU(2) gauge field takes the 
form\ct{KleKunBri92}
\ben
gW_idx^i = if_1(r)(-\sin\vph\si_1+\cos\vph\si_2)d\th + 
if_2(r)\si_3 d\vph,
\een
where $f_1$ and $f_2$ are functions which tend to 1 at infinity and 
vanish at the origin.
The magnetic charge $\tilde{Q}$ 
in the region $z\ge 0$ can be expressed as an 
integral of the magnetic field over a surface consisting of 
a hemisphere at infinity $N$ plus the equatorial plane $E$:
\ben
\tilde{Q}_{z\ge 0} = \int_{N+E} dx^i B_i^{\rm em}.
\een
Using the expression (\ref{eEMfNA}) for the electromagnetic field tensor, 
we see that the only contribution to this surface integral comes from 
the ``non-Abelian'' parts involving derivatives of 
the isovector field $\phh^a$.  On the equatorial plane, both 
$\pa_i\phh^a$ and  $\phh^aD_i\phh^bD_j\phh^c\ep^{abc}$ vanish.  Thus 
the magnetic charge $\tilde Q$ is given in terms of the integral
\ben
\tilde{Q}_{z\ge0} = 
{\sin\thw\over g}\int_N \phh^a\pa_i\phh^b\pa_j\phh^c\ep^{abc}.
\een
Recall that $\phh^a$ is constant on the boundary of the hemisphere: 
thus the hemisphere is effectively a 2-sphere, and integral on the 
right hand side of this equation measures the winding number of 
the isovector field around the unit sphere $\phh^2=1$.  Therefore, 
{\it for axisymmetric, 
parity-invariant configurations}, the magnetic 
charge in the region $z\ge 0$
is quantized in units of $4\pi\sin\thw/g$.
The same considerations apply to the region $z\le 0$.  The parity 
operation reverses magnetic fields, and so switches the sign of 
magnetic charge.  Thus the charge in the other hemisphere must be 
equal and opposite.

\sec{SPHALERONS AND ELECTROWEAK STRINGS}
The Standard Model contains another non-trivial classical solution, 
which is a vortex carrying Z-flux.\twoct{Nam77}{Vac92} It is 
not topologically stable, and is dynamically stable only in the 
unphysical region $\sin\thw\simeq 1$.\ct{JamPerVac93}  Because 
of their topological triviality, they can end, and Nambu\ct{Nam77} 
showed that they terminate on monopoles, which have magnetic 
charge $\pm 4\pi\sin\thw/g$.  Nambu proposed that there was a massive 
long-lived excitation in the Standard Model: 
a spinning segment of electroweak 
string, with oppositely charged monopoles at each end.
This ``dumb-bell'' bears some resemblance to the 
sphaleron,\twoct{HinJam94}{BarVacBuc94} which we saw in the last 
section also has a quantized 
magnetic dipole within it.  This section explores the 
connection between the two field configurations, amplifying some remarks 
made in Ref. {\ref{HinJam94}}.

\begin{figure}[bh]
\epsfxsize=3.4in
\centerline{\epsfbox{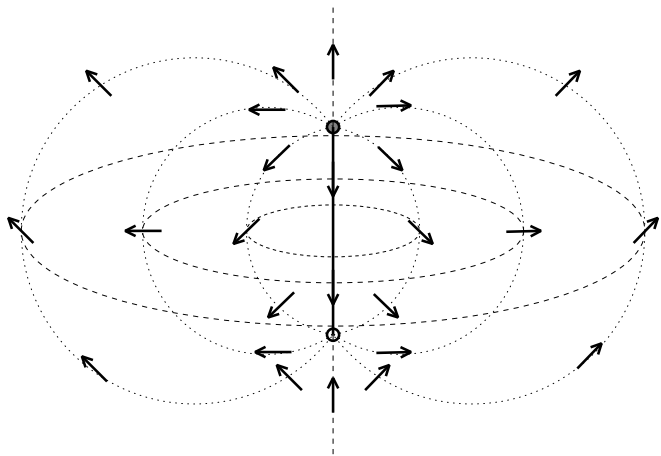}}
\caption{Isovector field configuration around Nambu' dumb-bell 
configuration.  The thick solid line is the line of zeroes 
of the Higgs field, dotted lines are lines of magnetic flux. 
Dashed lines guide the eye on the equatorial plane}
\end{figure}

The Higgs field in Nambu's dumb-bell configuration vanishes along a 
line segment, taking the form away from this line 
\ben
\Phi = \pmatrix{\cos\Th\cr\sin\Th e^{i\vph}\cr}{v\over\sqrt 2}, 
\qquad \cos2\Th = \cos\th_+ - \cos\th_- + 1,
\een
where $\th_\pm$ are polar angles measured from the ends of the line 
segment (see Figure 3).  The gauge fields can be written in terms 
of the Higgs field, by solving the equation  
$D_i\Phi=0$:
\bea
gW_i^a &=& -\ep^{abc}\phh^b\pa_i\phh^c - i \cos^2\thw \phh^a 
(\Phh^{\dag}\dbw_i\Phh), \\
g'Y_i  &=& -i\sin^2\thw (\Phh^{\dag}\dbw_i\Phh).
\eea
This solution actually requires two extra pieces of information: that 
vortices carry only Z flux (and no electromagnetic flux); and a gauge 
choice, which sets a possible arbitrary abelian part $(\al_i\phh^a, 
\al_i)$ to zero.  The second point is a crucial one, for the 
sphaleron is usually exhibited in the radial gauge, so direct comparisons 
can be misleading.  In Nambu's paper the fact that a gauge choice has 
been made is rather implicit:  a clearer discussion (for the 
Georgi-Glashow model) was given by Manton,\ct{Man77} who showed that 
in this gauge $\phh^a$ is constant along lines of magnetic flux, or 
\ben
B^{\rm em}_i \pa_i\phh^a = 0.
\een
This is illustrated in Figure 2,  which depicts 
the flux lines originating from the ends of the line segment.  An 
important point about this configuration is that away from the 
singular line the fields obey the equations of motion: that is, not 
only does the covariant derivative of the Higgs vanish, but also 
$\ep_{ijk}\pa_jB^{\rm em}_k = 0$.

Now, for the purposes of comparison, let us try and 
construct a sphaleron-type configuration in this gauge, which also 
vanishes on a line segment.  The most 
convenient way to do this is to use prolate spheroidal coordinates 
$(\mu,\be,\vph)$.\ct{MorFes53}  
The surfaces of constant $\mu$ are increasingly 
prolate as $\mu\to 0$, collapsing to a line of length $2d$ 
at $\mu=0$, while becoming 
more spherical as $\mu\to\infty$.  $(\be,\vph)$ are polar coordinates 
on the spheroids.  If $r_\pm$ are the distances from the ends of the 
line segment then $d\cosh\mu = (r_++r_-)/2$, and $\cos\be = 
(r_+-r_-)/2d$.  Then we may adapt (\ref{eSph}) by redefining $U$:
\ben
 U(\hat{\bxi}) = [\cos\chi(\mu) + 
 i\sin\chi(\mu)\bsi\cdot\hat{\bxi}]\si_2,
\een
where $\hat{\bxi} = 
(\sin\be\cos\vph,\sin\be\sin\vph,\cos\be)$; and $\chi(0)=\pi/2$, 
$\chi(\infty) = 0$.  In the limit that $d\to0$, $\hat{\bxi} \to 
\bxh$, and away from the origin the configuration is just a gauge 
transformation $\Om=\exp(i(\pi/2-\chi)\bsi\cdot\bxh)$ 
of the sphaleron.  The isovector field can 
be used to trace the magnetic flux lines:
\ben
\phh^a = \cos2\chi(\de^{a3} - \xih^a\xih^3) + \sin2\chi\ep^{a13} 
\xih^i + \xih^a\xih^3.
\een
\begin{figure}[h]
\epsfxsize=3.4in 
\centerline{\epsfbox{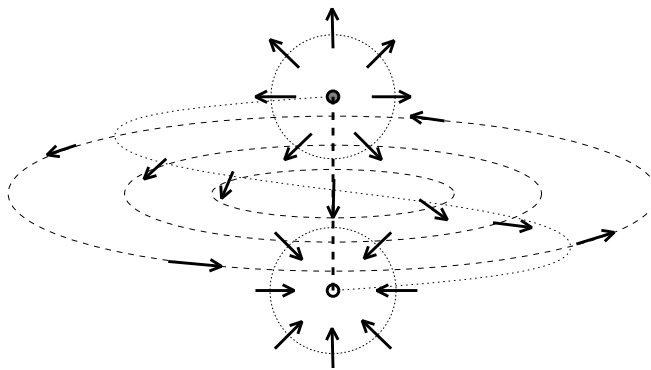}}
\caption{Isovector field configuration around the stretched sphaleron. 
Only one line of magnetic flux is shown.  Note the twist as it 
travels between the ends of the line of Higgs field zeroes.}
\end{figure}
Figure 3 shows the isovector field configuration around this 
``stretched'' sphaleron.  It is clear that the magnetic field has an 
azimuthal component: the lines of flux twist through $\pi$ as they 
travel from one pole to another.  This twist is the source of the 
difference between the sphaleron and the dumb-bell, for it turns 
out to be crucial in supplying the Chern-Simons number.\ct{VacFie94} 
Twisting the field is an unnatural thing to do, for the magnetic 
field cannot sustain such a twist without a current.  The 
current can only exist in the core of the string segment, where 
the Higgs field is off the vacuum and $F^a_{ij}D_j\phh^a$ does 
not vanish.  Thus, without an external electromagnetic current, 
the twist can exist only within the string.

The true significance of these field configurations is 
rather obscure, for they are not solutions of the field equations.  
However, they may be solutions to the equations subject to a constraint: 
that the Higgs field is forced to vanish on a line.  This 
is certainly a well-defined variational problem, for the constraint 
is gauge-invariant.  However, to find which (if either) of the dumb-bell 
or the stretched sphaleron is a solution seems exceedingly hard.

\def\ZP{Zeit. Phys.}
\def\JP{J. Phys.}
\def\NP{Nucl. Phys.}
\def\PL{Phys. Lett.}
\def\PR{Phys. Rev.}
\def\PRL{Phys. Rev. Lett.}
\def\ref #1 #2 #3 #4 #5 #6{#1, {\it #2} {\bf #3}, #4 (#5)#6\ }

\pagebreak[1]
\sec{REFERENCES}
\nobreak
\begin{enumerate}
\parskip 0pt
\item\label{HinJam94} M. Hindmarsh and M. James, {\it \PR\ D}, 
(to appear, 1994).
\item\label{tHo74} \ref{G. 't Hooft} {\NP} B79 276 1974 .
\item\label{Nam77} \ref{Y. Nambu} {\NP} B130 505 1977 .
\item\label{Vac93} T.Vachaspati, ``Electroweak Strings: 
a Progress Report'', {in}: proceedings
of Texas/\-PASCOS 92: Relativistic Astrophysics and Particle Cosmology, 
{\it Ann. N.~Y. Acad. Sci.\ } Vol. 688, (1993). 
\item\label{BarVacBuc94} M. Barriola, T. Vachaspati and M. Bucher, Embedded 
Defects, {\it \PR\ D} (to appear, 1994).
\item\label{KleKunBri92} \ref{B. Kleihaus, J. Kunz and Y. Brihaye} {\PR} 
{D46} {3587} 1992 .
\item\label{KliMan84} \ref{F. Klinkhamer and N. Manton} {\PR} {D30} {2212} 
1984 .
\item\label{Jam92} \ref{M. James} {\ZP} {C55} {515} 1992 .
\item\label{Pol74} \ref{A.M. Polyakov} {JETP Lett.} {20} {194} 1974 .
\item\label{Vac92} \ref{T. Vachaspati} {\PRL} {68} {1977} 1992 .
\item\label{JamPerVac93} \ref{M. James, L. Perivolaropoulos and 
T. Vachaspati} {\NP} {B395} {534} 1993 .
\item\label{Man77} \ref{N. Manton} {\NP} {B126} {525} 1977 .
\item\label{MorFes53} P. Morse and H. Feschbach, ``Methods of 
Mathmatical Physics,'' McGraw-Hill, New York (1953).
\item\label{VacFie94} T. Vachaspati and G.B. Field, {\it Tufts U. preprint} 
(1994).
\end{enumerate}

\end{document}